\documentclass[prc,unsortedaddress,preprint,amsmath,amsfonts,amssymb,showpacs,floatfix,nofootinbib]{revtex4-1}
\usepackage{epsfig}
\usepackage{multirow}
\usepackage{amsmath}
\usepackage{relsize}
\usepackage{longtable}
\usepackage{color}
\begin{document}

\title{Internal and external radiative widths in the combined R-matrix  and potential model formalism}

\author{A. M. Mukhamedzhanov$^{1}$}
\email{akram@comp.tamu.edu} 
\author{Shubhchintak$^{2}$}
\email{shub.shubhchintak@tamuc.edu} 
\author{C. A.  Bertulani$^{2,3}$}
\email{carlos.bertulani@tamuc.edu} 
\author{T. V. Nhan Hao$^{2}$}
\email{hao.tran@tamuc.edu} 

\affiliation{$^{1}$Cyclotron Institute, Texas A$\&$M University, College Station, TX 77843, USA}
\affiliation{$^{2}$Department of Physics and Astronomy, Texas A$\&$M University-Commerce, Commerce, TX 75429, USA}
\affiliation{$^{3}$Department of Physics and Astronomy, Texas A$\&$M University, College Station, TX 77843, USA}

\date{\today}

\begin{abstract}
Using the $R$-matrix approach we calculate the radiative width for a resonance decaying to a bound state through electric dipole, $E1$, transitions. The total radiative width is determined by the interference of the nuclear internal and external radiative width amplitudes. For a given channel radius the external radiative width amplitude is model independent and is determined by the asymptotic normalization coefficient (ANC) of the bound state to which the resonance decays. It also depends on the partial resonance width. To calculate the internal radiative width amplitude we show that a single-particle potential model is appropriate.  We compare our results with a few experimental data.      
\end{abstract}
\pacs{21.10.Jx, 25.40.Ny, 23.20.Lv, 27.20.+n}

\maketitle

\section {Introduction}
In nuclear astrophysics several important nucleon capture reactions take place through resonance states which then decay to  bound states. The total capture cross section for such reactions is then given by the interference of resonant and non-resonant contributions. Many theoretical models for resonant and non-resonant cross sections require proper knowledge of the initial and final state and the nature and multipolarity of the transition \cite{lane_thomas,holt78,barker91,huang2010}. In addition, the resonant cross section can also be expressed in terms of the radiative width and the partial width of the resonance \cite{lane_thomas,holt78,barker91}. In fact, the radiative width is one of the important observables whose precise value is required in order to determine the resonance capture cross sections accurately. 

The radiative width amplitude in terms of the initial ($\Psi$) continuum and final ($\phi$) bound state wave functions  can be written as $\langle\phi|\hat{O}|\psi\rangle$, with $\hat{O}$ being the electromagnetic operator.  To calculate it the $R$-matrix approach is often used  \cite{lane_thomas,holt78,barker91,akram_11,akram_pang15}. 
In the $R$-matrix approach the radiative width amplitude is given by the sum of the nuclear internal and external (channel) parts. The channel radiative width amplitude depends only on one model parameter, namely, the channel radius, and for a given channel radius the channel radiative width amplitude is model-independent. Apart from this, to calculate the channel radiative width amplitude one needs to know two observables:  the ANC of the final bound state and partial resonance width. Therefore, with precise knowledge of these quantities, the channel radiative width amplitude can be calculated quite accurately. The channel radiative width amplitude is a complex quantity and its imaginary part puts a lower limit on the radiative width \cite{akram_pang15}. Contrary to this, the internal radiative width amplitude is a real and model-dependent quantity. In the $R$-matrix method the internal radiative width amplitude is usually taken as a fitting parameter.

In this paper we use the single-level $R$-matrix approach along with the single-particle potential model to calculate the radiative width, where the internal radiative width amplitude and its relative sign to that of the external radiative width amplitude are calculated using a potential model consistently \cite{radcap}. This work  follows the $R$-matrix formalism presented in \cite{holt78,barker91,akram_pang15}.  The radiative width amplitude is split into two parts, internal and external . The internal radiative width amplitude and the relative sign of the internal and external radiative width amplitudes were fitting parameters. In \cite{akram_pang15} the calculated external width amplitude was used to  to set a lower limit on the radiative width amplitude. 
Our work presents further development of the $R$-matrix formalism developed in \cite{holt78,barker91,akram_pang15} by calculating the internal width amplitude and its sign rather then using them as fitting parameters. After calculating the internal width amplitude we calculate also the total radiative width.
We consider both the decay of the resonances to bound states and decay of the subthreshold resonance to the bound state important for nuclear astrophysics.

This paper is organized in the following way. In section II, we describe our formalism to calculate the internal and external radiative width amplitudes and the total radiative width. In section III we discuss few practical cases and present our calculated radiative widths for those. Our conclusions are presented in section IV. 

\section{Formalism}
\label{sec:2}
We consider the radiative capture reaction $x + A \rightarrow B^*\rightarrow B + \gamma$, where the intermediate resonance $B^*$ decays to final bound state $B = (xA)$. We define $\Psi_{i}^{(+)}$ as the initial scattering wave function and $\phi_{B}$ as the final bound-state wave function. Let $R$ be the channel radius, which divides the internal and external regions of the resonance system $(x+A=B^*)$ having relative momentum $k$ in the initial state. For very low initial energies ($1/k >> R$), one can use the long-wavelength approximation allowing one to treat the individual particles as structureless. Then the initial scattering wave function can be written as,
\begin{eqnarray}
\Psi_{i}^{(+)}=\phi_{x}\phi_{A}\Psi_{l_{i}}^{(+)} \label{a1},
\end{eqnarray}
where $\phi_{x}$ and $\phi_{A}$ are the internal state wave functions of nuclei $x$ and $A$, respectively, and $\Psi_{l_{i}}^{(+)}$ is the scattering wave function in the partial wave $l_{i}$. In the long-wavelength approximation, one can write the reaction amplitude as \cite{akram_pang15}
\begin{eqnarray}
M &=& \sqrt{\frac{8\pi(L+1)}{L}}\frac{k_{\gamma}^{L+1/2}}{(2L+1)!!\sqrt{2J_f+1}} \langle \phi_B|\hat{O}_L|\phi_{x}\phi_{A}\Psi_{l_{i}}^{(+)}\rangle\nonumber\\
&=&\sqrt{\frac{8\pi(L+1)}{L}}\frac{k_{\gamma}^{L+1/2}}{(2L+1)!!\sqrt{2J_f+1}} \langle I^{B}_{xA}|\hat{O}_L|\Psi_{l_{i}}^{(+)}\rangle.
\label{a2}
\end{eqnarray}  
The integration in Eq. (\ref{a2}) is performed over $r$. $\,\hat{O}_L$ is the electromagnetic transition operator of multipolarity $L$, which in the long-wavelength approximation depends only on the distance $r$ between the center of mass of $x$ and of $A$,  $\,k_\gamma$ is the wave number of the photon and $J_f$ is the spin of the final bound state. 
$\,I^{B}_{xA}$ is the radial overlap function of the bound state of nuclei $\,x$, $\,A$ and $\,B$  given by $\,I^{B}_{xA} =\langle \phi_{x}\phi_{A}|\phi_B\rangle$ where the integration is performed over the internal coordinates of nuclei $x$ and $A$. Hence $I^{B}_{xA}$  depends only on $r$. 
 
Following the $R$-matrix formalism we split the scattering wave function into two parts: the internal ($\Psi_{l_{i}\,int}^{(+)}$, for $r\leq R$) and external  ($\Psi_{l_{i}\,ext}^{(+)}$, for $r\geq R$). Because of the linear dependence of the total radiative capture amplitude on $\Psi_{l_{i}}^{(+)}$, we can write it as the sum of the internal and external radiative capture amplitudes,
\begin{eqnarray}
M = M_{int}+M_{ext},  
\label{a3}
\end{eqnarray}
where
\begin{eqnarray}
M_{int} &=& \sqrt{\frac{8\pi(L+1)}{L}}\frac{k_{\gamma}^{L+1/2}}{(2L+1)!!\sqrt{2J_f+1}}
\langle I^{B}_{xA}|\hat{O}_L|\Psi_{l_{i}\,int}^{(+)}\rangle \Big|_{r\leq R}
\label{a4}
\end{eqnarray}
and
\begin{eqnarray}
M_{ext} &=& \sqrt{\frac{8\pi(L+1)}{L}}\frac{k_{\gamma}^{L+1/2}}{(2L+1)!!\sqrt{2J_f+1}}
 \langle I^{B}_{xA}|\hat{O}_L|\Psi_{l_{i}\,ext}^{(+)}\rangle \Big|_{r\geq R}.  
\label{a5}
\end{eqnarray}
It is clear that  $M_{int}$ is contributed by the radial integral taken over the nuclear interior ($r\leq R$) whereas $M_{ext}$ is contributed by the radial integral over the nuclear exterior ($r\geq R$). 

In the single-channel R-matrix method, the internal wave function for the case of an isolated narrow resonance is given for $\,r\leq R$   by \cite{holt78}
\begin{eqnarray}
\Psi_{l_{i}\,int}^{(+)} (k,r) = - i e^{-i\delta_{l_i}^{hs}} \frac{[\Gamma_{J_{i}}^{(0)}]^{1/2}}{E_R-E-i\frac{\Gamma_{J_{i}}^{(0)}}{2}} X_{int}(k,r),
\label{a6}
\end{eqnarray}
where  $\,\delta_{l_i}^{hs}$ is the hard sphere scattering phase shift for the partial wave $l_i$, $\,E = k^{2}/2\,\mu$ is the $x-A$ relative kinetic energy,  $\mu$ is their  reduced mass,   $\,E_R$ is the real part of the resonance energy and $\,X_{int}$ is the real internal $R$-matrix wave function of the level closest to the resonance\footnote[1]{In this paper we use the single-level $R$-matrix approach.}.
$\Gamma_{J_{i}}^{(0)}$ is the observed partial width of the resonance having spin $J_i$ for the decay to the channel $x + A$. It is related to the observed reduced width $(\gamma_{J_{i}}^{(0)})^{2}$ as 
\begin{eqnarray}
\Gamma_{J_{i}}^{(0)}= 2\,P_{l_{i}}(E,R)\,(\gamma_{J_{i}}^{(0)})^{2},
\label{Gammai0gammai01}
\end{eqnarray}
where $P_{l_{i}}(E,R)$ is the penetrability factor and $l_{i}$ is the angular orbital moment of the resonance in the channel $x + A$. The observed reduced width  is given by
\begin{eqnarray}
(\gamma_{J_{i}}^{(0)})^{2}= (\gamma_{J_{i}})^{2}\,N_{i}
\label{gammai0gammai1}
\end{eqnarray}
with \cite{lane_thomas,barker80}
\begin{eqnarray}
N_{i}= \frac{1}{1- \frac{{\rm d} \Delta_{l_{i}}(E, R)}{{\rm d}E}\,\Big|_{E=E_{R} } }
\label{Ni1}
\end{eqnarray}
and  
\begin{eqnarray}
\Delta_{l_{i}}(E,R)= - k\,R\,\frac{F_{l_i}'(k,r)\,F_{l_i}(k,r)  + G_{l_i}'(k,r)\,G_{l_i}(k,r)}{F_{l_i}^2(k,r)  +  G_{l_i}^2(k,r)  }\,\Big|_{r=R}\,\gamma_{J_{i}}^{2}\,,
\label{Delta1}
\end{eqnarray}
where $\gamma_{J_{i}}$ is the $R$-matrix formal reduced width amplitude,   $F_{l_i}$ and $G_{l_i}$ are the regular and singular Coulomb solutions. The prime stands for the differentiation over $\rho= k\,r$. 

In this work we calculate the internal $R$-matrix wave function $\,X_{int}$ using the potential model \cite{radcap}, where the scattering potential is adjusted to reproduce the resonance at the right position. The wave function $X_{int}$  is normalized to unity over the internal region: 
\begin{eqnarray}
\int_0^{R} dr~X_{int}^2(k,r) = 1.
 \label{a7}
\end{eqnarray}

The scattering wave function in the external region  ($ r\geq R$)   is given by
\begin{eqnarray}
\Psi_{l_{i}\,ext}^{(+)}(k,r)=\sqrt{\frac{1}{v}}[I_{l_i}(k,r)-S_{l_{i}}\,O_{l_i}(k,r)], 
\label{a8}
\end{eqnarray}
where $v=k/\mu$ is the $x-A$ relative velocity, $\,I_{l_{i}}$ and $\,O_{l_{i}}$ are the incoming and outgoing spherical waves in the partial wave $\,l_{i}$ and $\,S_{l_{i}}$ is the elastic scattering $\,S$-matrix element. The elastic scattering $S$-matrix element is given by
\begin{eqnarray}
S_{l_{i}}= e^{-2i\delta_{l_i}^{hs}}\Bigg(1   +  \frac{i\Gamma_{J_{i}}^{(0)}}{E_R-E-i\frac{\Gamma_{J_{i}}^{(0)}}{2}}\Bigg). 
\label{a9}
\end{eqnarray}

In the R-matrix approach  the hard-sphere scattering phase shift ($\delta_{l_i}^{hs}$) is given by
\begin{eqnarray}
e^{-2i\delta_{l_i}^{hs}} = \frac{I_{l_i}(k,R)}{O_{l_i}(k,R)}. 
\label{a10}
\end{eqnarray}

Using Eq. (\ref{a9}) we can rewrite the external wave function $\Psi_{l_{i}\,ext}^{(+)}(k,r)$ as
\begin{align}
\Psi_{l_{i}\,ext}^{(+)}(k,r) = \Psi_{l_{i}\,ext}^{(+)(NR)}(k,r) + \Psi_{l_{i}\,ext}^{(+)(R)}(k,r).
\label{extNRR1}
\end{align}
Here 
\begin{align}
 \Psi_{l_{i}\,ext}^{(+)(NR)}(k,r)= \sqrt{\frac{1}{v}}[I_{l_i}(k,r)-  e^{-2i\delta_{l_i}^{hs}}\,O_{l_i}(k,r)]
\label{extNRwf1}
\end{align}
is the external wave function contributing to the non-resonant radiative capture in the $R$-matrix approach
and 
\begin{eqnarray}
\Psi_{l_{i}\,ext}^{(+)(R)}(k,r)=-\sqrt{\frac{1}{v}} ~ \frac{i\Gamma_{J_{i}}^{(0)}}{E_R-E-i\frac{\Gamma_{J_{i}}^{(0)}}{2}} e^{-2i\delta_{l_i}^{hs}} O_{l_i}(k,r),  
\label{a11}
\end{eqnarray}
is the external wave function contributing to the resonant capture in the external region.

Correspondingly,  the external radiative capture amplitude $M_{ext}$ can be split into two parts:
\begin{align}
M_{ext} =  M^{NR} + M_{ext}^{(R)},
\label{MextNRR1}
\end{align}
where 
\begin{align}
 M^{NR}= \sqrt{\frac{8\pi(L+1)}{L}}\frac{k_{\gamma}^{L+1/2}}{(2L+1)!!\sqrt{2J_f+1}}
 \langle I^{B}_{xA}|\hat{O}_L|\Psi_{l_{i}\,ext}^{(+)(NR)}\rangle \Big|_{r\geq R} 
\label{NRampl1}
\end{align}
is the non-resonant radiative capture amplitude in the $R$-matrix approach and 
\begin{eqnarray}
M_{ext}^{(R)} &=& \sqrt{\frac{8\pi(L+1)}{L}}\frac{k_{\gamma}^{L+1/2}}{(2L+1)!!\sqrt{2J_f+1}}
 \langle I^{B}_{xA}|\hat{O}_L|\Psi_{l_{i}\,ext}^{(+)(R)}\rangle \Big|_{r\geq R}.  
\label{a5}
\end{eqnarray}
is the external radiative capture amplitude in the $R$-matrix approach. 
Then, in the $R$-matrix approach, we can write the radiative capture amplitude as
\begin{align}
M= M^{(R)}  + M^{(NR)}.
\label{MRNR1}
\end{align}
It is worth noting that in the $R$-matrix method the radiative capture amplitude is split into resonant part, which is contributed 
by both internal and external amplitudes and the non-resonant amplitude, which is entirely contributed only by the external non-resonant radiative capture while the internal non-resonant radiative capture is absorbed into the internal resonant capture. 

The resonant radiative capture amplitude is
\begin{align}
&M^{(R)}= M_{int} + M_{ext}^{(R)}= \sqrt{\frac{8\pi(L+1)}{L}}\frac{k_{\gamma}^{L+1/2}}{(2L+1)!!\sqrt{2J_f+1}}
\langle I^{B}_{xA}|\hat{O}_L|\Psi_{l_{i}\,int}^{(+)}\rangle \Big|_{r\leq R}                                           \nonumber\\
& +  \sqrt{\frac{8\pi(L+1)}{L}}\frac{k_{\gamma}^{L+1/2}}{(2L+1)!!\sqrt{2J_f+1}}
 \langle I^{B}_{xA}|\hat{O}_L|\Psi_{l_{i}\,ext}^{(+)}\rangle \Big|_{r\geq R}.
\label{Rampl1}
\end{align}


Matching the internal and external wave functions at the channel radius $R$ we get  at  $\,E = E_R$
\begin{eqnarray}
X_{int}(k_{R},R)= \sqrt{\frac{\mu}{k_{R}}} \sqrt{\Gamma_{J_{i}} } \sqrt{F_{l_i}^2(k_{R},R) + G_{l_i}^2(k_{R},R)} =\sqrt{2\,\mu\,R}\,\gamma_{J_{i}}^{(0)} ,
\label{a12}
\end{eqnarray}
where $E_{R}= k^{2}_{R}/(2\,\mu)$,  $\,\Gamma_{J_{i}}^{(0)}= 2\,P_{l_{i}}(k,R)\,(\gamma_{J_{i}}^{(0)})^{2}$, $\,\gamma_{J_{i}}^{(0)}$ is the observed  reduced width amplitude in the $R$-matrix approach, $P_{l_{i}}(k,R)= 2\,k\,R/\big(F_{l_i}^2(k,R)  +  G_{l_i}^2(k,R)\big )$ is the penetrability factor.
Thus early introduced  Eq.  (\ref{a6}) provides the correct $R$-matrix definition of $X_{int}$ at  $r=R$ in terms of the reduced width amplitude, see Eq. (iV.1.10a) \cite{lane_thomas}.

Using Eqs. (\ref{a6}) and (\ref{a11}) we get from Eqs. (\ref{a4}) and (\ref{a5}) the internal and external radiative capture amplitudes:
\begin{eqnarray}
M_{int}=- i e^{-i\delta_{l_i}^{hs}} \frac{\sqrt{\Gamma_{J_{i}}^{(0)}} \gamma_{\gamma J_f}^{J_i}(int)}{E_R-E-i\frac{\Gamma_{J_{i}}^{(0)}}{2}}, \label{a13}
\end{eqnarray}
and
\begin{eqnarray}
M_{ext}= -i e^{-i\delta_{l_i}^{hs}} \frac{\sqrt{\Gamma_{J_{i}}^{(0)}} \gamma_{\gamma J_f}^{J_i}(ch)}{E_R-E-i\frac{\Gamma_{J_{i}}^{(0)}}{2}} + M_{nr}, \label{a14}
\end{eqnarray}
where  $M_{nr}$ is the external part of non-resonant (direct) radiative capture amplitude. The internal part of the direct radiative capture amplitude is absorbed in $M_{int}$. In the above equations quantities $\gamma_{\gamma J_f}^{J_i}(int)$ and $\gamma_{\gamma J_f}^{J_i}(ch)$ are the internal and external (channel) radiative width amplitudes for the decay of resonance with spin $J_i$ to the bound state having spin $J_f$. They are given by
\begin{eqnarray}
\gamma_{\gamma J_f}^{J_i}(int) &=& \sqrt{\frac{8\pi(L+1)}{L}}\frac{k_{\gamma}^{L+1/2}}{(2L+1)!!\sqrt{2J_f+1}}
\langle I^{B}_{xA} (r)|\hat{O}_L|X_{int}(r)\rangle \Big|_{r\leq R}
\label{a15}
\end{eqnarray}
and
\begin{eqnarray}
\gamma_{\gamma J_f}^{J_i}(ch) &=& \sqrt{\frac{8\pi(L+1)}{L}}\frac{k_{\gamma}^{L+1/2}}{(2L+1)!!\sqrt{2J_f+1}}
\sqrt{\frac{\mu}{k}\,\Gamma_{J_{i}}^{(0)}}\langle I^{B}_{xA}(r)|\hat{O}_L|e^{-i\delta_{l_i}^{hs}}O_{l_i}(r)\rangle \Big|_{r\geq R}.  \label{a16}
\end{eqnarray}

In the potential model, the radial overlap function $I^{B}_{xA}(r)$ can be expressed in terms of the bound state wave function as,
\begin{eqnarray}
I^{B}_{xA}(r)=\sqrt{S_{l_f J_f I}} ~\phi_{l_f J_f I}^B(r), \label{a17}
\end{eqnarray}
where $S_{l_f J_f I}$ is the spectroscopic factor of the final bound state with $l_f$ being the  $x-A$ relative angular momentum of the bound state  and $I$ is the channel spin.  The tail of the bound-state wave function behaves as
\begin{eqnarray}
\phi_{l_f J_f I}^B(r) \stackrel{r> R}{\approx} b_{l_f J_f I} ~W_{-\eta_f, l_f+1/2}(2\kappa_{f} r), \label{a18}
\end{eqnarray}
where $W_{-\eta_f, l_f+1/2}(2\kappa_{f} r)$ is the Whittaker function, $\kappa_{f}$ is the bound-state wave number  and $\,\eta_f$ is the  Coulomb parameter of the bound state. $b_{l_f J_f I}$ is the single-particle ANC and its value depends upon the bound-state potential. Therefore, in the external region the overlap function becomes
\begin{eqnarray}
I^{B}_{xA}(r)= C_{l_f J_f I}~~ W_{-\eta_f, l_f+1/2}(2\kappa_{f} r),
 \label{a19}
\end{eqnarray}
where 
\begin{eqnarray}
C_{l_f J_f I} = b_{l_f J_f I}~\sqrt{S_{l_f J_f I}} 
\label{ANCsnpANC1}
\end{eqnarray}
is the ANC of the final bound state. 

Now we will discuss the expressions for the internal and channel radiative width amplitudes,  which correspond to both ``resonance $\to$ bound state" and ``subthreshold resonance $\to $  bound state" transitions. Note that for the transition  ``subthreshold resonance $\to$ bound state" the resonance energy is negative: $E_{R}=-\epsilon_{i}$, where $\epsilon_{i}$ is the binding energy of the subthreshold state. For nuclear astrophysical application  we are interested in the radiative capture cross sections at $E \to 0$. 

Following Refs. \cite{barker80,barker91,akram_pang15}, the internal and channel radiative width amplitudes for ``the resonance to the bound-state" transitions at the resonance energy ($E= E_R$) are then simplified to (in MeV and fm units)
\begin{eqnarray}
\gamma_{\gamma J_f}^{J_i}(int) &=& \sqrt{S_{l_f J_f I}} \sqrt{\frac{\lambda_N . 931.5}{137 E}} (R k_{\gamma})^{L+\frac{1}{2}} \mu^L 
\Bigg(\frac{Z_x}{m_x^L}+(-1)^L \frac{Z_A}{m_A^L}\Bigg)  \sqrt{\frac{(L+1)(2L+1)}{L}} \nonumber\\
&\times&\frac{1}{(2L+1)!!}\,\sqrt{k R} \sqrt{(2l_i+1)(2J_f+1)}
(-1)^{L+l_f+I+J_i}~ C^{l_f 0}_{l_i 0 L 0} ~\left\{\begin{array}{ccc} L & l_f & l_i\\ I & J_i & J_f\end{array}\right\} \frac{1}{R^{L+1}}\nonumber\\
& \times& \sqrt{\frac{k \hbar^2 }{\mu}}\, \int_0^{R} dr r^L \phi_{l_f J_f I}^B(r) X_{int}(k,r)\,N_{i}^{1/2}
\label{a20}
\end{eqnarray}
and 
\begin{eqnarray}
\gamma_{\gamma J_f}^{J_i}(ch) &=& C_{l_f J_f I} \sqrt{\frac{\lambda_N . 931.5}{137 E}} (R k_{\gamma})^{L+\frac{1}{2}} \mu^L 
\Bigg(\frac{Z_x}{m_x^L}+(-1)^L \frac{Z_A}{m_A^L}\Bigg)  \sqrt{\frac{(L+1)(2L+1)}{L}} \nonumber\\
&\times&\frac{1}{(2L+1)!!} \sqrt{\Gamma_{J_{i}}^{(0)}}~ \sqrt{k R} \sqrt{(2l_i+1)(2J_f+1)}
(-1)^{L+l_f+I+J_i}~ C^{l_f 0}_{l_i 0 L 0}~ \left\{\begin{array}{ccc} L & l_f & l_i\\ I & J_i & J_f\end{array}\right\}\frac{1}{R^{L+1}}\nonumber\\
& \times& \int_{R}^{\infty} dr r^L W_{-\eta_f^{bs}, l_f+1/2}(2\kappa r) e^{-i\delta^{hs}}O_{l_i}(k,r), \nonumber\\
~~\label{a21}
\end{eqnarray}
where $\lambda_N=0.2118 $ fm is the nucleon Compton wavelength, $Z_i$ and $m_i$ are the charge and mass of  particle $\,i$, $\,C^{l_f 0}_{l_i 0 L 0}$ is the Clebsch-Gordan coefficient and the quantity in curly bracket is the 6-j symbol. Note that the above radiative width amplitudes are expressed in  MeV$^{1/2}$. All masses are expressed in units of MeV$/c^2$,  $\,E$ and $\,\Gamma_{J_{i}}^{(0)}$ are in  MeV and the wave number in fm$^{-1}$. 
The bound-state wave function $\phi_{l_f J_f I}^B(r)$ in Eq. (\ref{a20}) is normalized to unity over the whole radial space ($0 \leq r < \infty$) and is calculated by solving the Schr{\"o}dinger equation with a Woods-Saxon (WS) potential, whose parameters are adjusted to get the corresponding binding energy of the state. The resonance scattering wave function in the internal region is given by $X_{int}$, which is normalized to unity over the internal region. The channel radiative width amplitude is proportional to $\sqrt{\Gamma_{J_{i}}^{(0)}}$. 

For the calculations of internal radiative width amplitude in the case of ``subthreshold to the bound-state transition" the factor $N_{i}^{1/2}$ should be dropped. 
The resonance width of the subthreshold resonance  is given by \cite{muk99} 
\begin{eqnarray}
\Gamma_{l_iJ_iI}^{(0)} = \frac{1}{\mu} P_{l_i}(E,R)\frac{[W_{-\eta^{bs}_{i},l_i+\frac{1}{2}}(2\kappa_i R)]^2}{R}(C_{l_iJ_iI})^2, 
\label{a28}
\end{eqnarray}
where $l_i$, $J_i$ and $I$ are the orbital angular momentum, spin and channel spin of the subthreshold state, respectively. $\kappa_i$ and $\eta_{i}^{bs}$ are the bound state wave number and Coulomb parameter of the subthreshold bound state. $C_{l_iJ_iI}$ is the ANC of the subthreshold bound state.    

It is clear that the internal radiative width amplitude is real because it involves the product of the real wave functions  $\phi_{l_f J_f I}^B(r)$ and $X_{int}$. On the other hand, the channel radiative width amplitude contains the complex function $e^{-i\delta^{hs}}O_{l_i}(k,r)$ and therefore is a complex quantity. Furthermore, the channel radiative width amplitude has only one model dependent parameter, which is the channel radius, whereas the internal radiative width is model dependent. 

Once the $\gamma_{\gamma J_f}^{J_i}(int)$ and $\gamma_{\gamma J_f}^{J_i}(ch)$ are calculated from Eqs. (\ref{a20}) and (\ref{a21}), we can find the total radiative width amplitude:
\begin{eqnarray}
\gamma_{\gamma J_f}^{J_i} = \gamma_{\gamma J_f}^{J_i}(int) + \gamma_{\gamma J_f}^{J_i}(ch).
 \label{a22}
\end{eqnarray}

The total radiative width $\Gamma_{\gamma J_f}^{J_i}$ is  given by the modulus square of the total radiative width amplitude,
\begin{eqnarray}
\Gamma_{\gamma J_f}^{J_i}=\Big|\gamma_{\gamma J_f}^{J_i}\Big|^2=\Big|\gamma_{\gamma J_f}^{J_i}(int) + \gamma_{\gamma J_f}^{J_i}(ch)\Big|^2, \label{a23}
\end{eqnarray}
which further can be written as
\begin{eqnarray}
\Gamma_{\gamma J_f}^{J_i}=\Big|\gamma_{\gamma J_f}^{J_i}(int) +  {\text Re}\Big[\gamma_{\gamma J_f}^{J_i}(ch)\Big]\Big|^2 + \Big({\text Im}\Big[\gamma_{\gamma J_f}^{J_i}(ch)\Big]\Big)^2.\nonumber\\ 
~~\label{a24}
\end{eqnarray}

${\text Re}\Big[\gamma_{\gamma J_f}^{J_i}(ch)\Big]$ (real part) and $\gamma_{\gamma J_f}^{J_i}(int)$ can interfere either constructively or destructively. Therefore, the imaginary part of the channel radiative width amplitude ${\text Im}\Big[\gamma_{\gamma J_f}^{J_i}(ch)\Big]$ gives the lower limit of the radiative width \cite{akram_pang15}. 

In the above equations we derived the radiative width at resonance energy, however, one can calculate it at any positive energy using the energy dependent relations for the partial resonance width and radiative width:
\begin{eqnarray}
\Gamma_{J_{i}}^{(0)}(E)=\frac{P_{l_i}(E)}{P_{l_i}(E_R)}\Gamma_{J_{i}}^{(0)}(E_R)
 \label{a25}
\end{eqnarray}
and
\begin{eqnarray}
\Gamma_{\gamma J_f}^{J_i}(E) = \Bigg(\frac{E+\epsilon_{f}}{E_R+\epsilon_{f}}\Bigg)^{2L+1}\Gamma_{\gamma J_f}^{J_i}(E_R),
 \label{a27}
\end{eqnarray}
where  $P_{l_i}(E)$ is the barrier penetrability  given by 
\begin{eqnarray}
P_{l_i}(E)=\frac{kR}{F_{l_i}^2(k,R)+G_{l_i}^2(k,R)}
\label{a26}
\end{eqnarray}
and $\epsilon_{f}$ is the binding energy of the state to which resonance decays.

For the decay of the subthreshold resonance to the lower lying bound state, the energy dependence of the radiative width  is given by
\begin{eqnarray}
\Gamma_{\gamma J_f}^{J_i}(E) = \Bigg(\frac{E+\epsilon_{f}}{\epsilon_{f}- \epsilon_{i}}\Bigg)^{2L+1}\Gamma_{\gamma J_f}^{J_i}(-\epsilon_{i}).
 \label{as27}
\end{eqnarray}
Using  Eq. (\ref{a24})  we can find $\Gamma_{\gamma J_f}^{J_i}(E) $ at $E>0$ and then from Eq. (\ref{as27})  the radiative width at the subthreshold bound state 
$\Gamma_{\gamma J_f}^{J_i}(-\epsilon_{i})$ can be easily calculated. 

\section{Results and discussions}
Using the formalism presented in the previous section, we now calculate the radiative width amplitudes (internal and external) for some cases which involve E1 decay of the resonance to the bound state. The calculated radiative widths are compared with the corresponding experimental  values. 
As we are using the $R$-matrix approach, the channel radius $R$ is a model parameter. Usually the channel radius is determined by using the relation $R = 1.4(x^{1/3}+A^{1/3})$, unless the experimental data for astrophysical factors are available and in those cases it is determined by fitting the experimental data. Here, $x$ means the mass number of a (valence) particle and $A$ is that of the nucleus. To calculate the external radiative width amplitude we use the experimental values of the partial resonance width and ANC of the bound state. Let us consider some particular cases.

1. {\it Decay of} $^{13}$N($\frac{1}{2}^+, E_x=2.365$ {\it MeV}) $\rightarrow$ $^{13}$N($\frac{1}{2}^-, E_x=0$ {\it MeV).}\\
We consider the decay of the $1/2^+$ resonance in $^{13}$N at $E_{R}=0.421$ MeV (where $E_{R}$ is the $p + {}^{12}{\rm C}$ resonance relative kinetic energy)  to the ground state $1/2^-$, having proton binding energy $\epsilon_{f}= 1.944$ MeV. This transition plays an important role in the radiative proton capture $^{12}$C + $p$ $\rightarrow$ $^{13}$N + $\gamma$ reaction, which is the very first reaction of the CNO cycle responsible for the energy generation in massive stars \cite{bbff_57}. 

The $1/2^+$ and $1/2^-$ states of $^{13}$N are obtained by coupling the $^{12}$C ($0^+$) core with 2$s_{1/2}$ and 1$p_{1/2}$ protons, respectively. We use the experimental ANC for the ground state of $^{13}$N,  $C_{l_{f}=1\,J_{f}=1/2\,I=1/2}=1.43 \pm 0.09$ fm$^{-1/2}$ \cite{yrmuk97}, and the proton resonance width $31.7 \pm 0.8$ keV \cite{Ajzenberg91}. Using Eq. (\ref{a21}), the channel radiative width amplitude  calculated for $R =4.6$ fm  and for the channel spin $\,I =1/2$ is $\,\gamma_{\gamma 1/2}^{1/2}(ch)=-0.519-i~0.018$ eV$^{1/2}$.   
The determination of $\,\gamma_{\gamma 1/2}^{1/2}(int)$ given by Eq. (\ref{a20}) requires the calculation of the bound-state wave function of the ground state and resonance wave function in the interior region ($r\leq R$).  We adopt the Woods-Saxon potential with geometry $r_0=1.25$ fm and $a=0.65$ fm and the depth of the spin-orbit potential  $-10$ MeV. With these potential parameters, the potential depths required in order to reproduce the ground and resonance state energies are $V_b =-43.525$ MeV and $V_c =-55.90$ MeV, respectively (the index $b$ stands for ``bound" and $c$ for ``continuum"). The single particle ANC is $b_{l_{f}=1\, J_{f}=1/2\, I=1/2}=2.008$ fm$^{-1/2}$. Then from  Eq. (\ref{ANCsnpANC1}) we get that the spectroscopic factor for the ground state is $0.51$.
These values yield the internal radiative width amplitude  obtained from Eq. (\ref{a20}) as $\gamma_{\gamma 1/2}^{1/2}(int)= - 0.262$ eV$^{1/2}$. Thus, when calculating the total radiative width the interference between the internal radiative width amplitude and the real part of the channel radiative width amplitude is constructive. 

The total radiative width calculated using Eq. (\ref{a24})  is $\Gamma_{\gamma 1/2}^{1/2} = 0.61 \pm 0.05$ eV, which is close to the previously measured radiative widths $0.65 \pm 0.07$ \cite{burtebaev}, 0.67 \cite{ajz70}, $0.50 \pm 0.04$ \cite{Ajzenberg91} and $0.45 \pm 0.05$ \cite{riess68}. We also checked that with 6.5 $\%$ variation in the channel radius, the radiative width changes only by 4 $\%$. To calculate the uncertainty $\Delta$ of the total radiative width  we use the equation
\begin{align}
\Delta = \sqrt{\Delta_{ANC}^{2} + \Delta_{\Gamma}^{2} + \Delta{R}^{2}},
\label{Delta1}
\end{align}
where $\Delta_{ANC}$,  $\,\Delta_{\Gamma}$  and  $\Delta{R}$ are the uncertainties of the radiative width caused by the uncertainty of the  experimental ANC,   of the partial resonance width  and the channel radius, correspondingly. Here, we assigned $10\%$ uncertainty for the square of the ANC and in all the cases below.

2. {\it Decay of} $^{13}$O($\frac{1}{2}^+$, $E_x=2.69$ {\it MeV}) $\rightarrow$ $^{13}$O($\frac{3}{2}^-$, $E_x=0$ {\it MeV}).\\
We now consider the transition of $1/2^+$ resonance at $E_{R} = 1.17$ MeV in $^{13}$O to the ground state $3/2^-$ with $\epsilon_{f}=1.515$ MeV. The $1/2^+$ and $3/2^-$ states of $^{13}$O are obtained by coupling the $^{12}$N ($1^+$) core with 2$s_{1/2}$ and 1$p_{1/2}$ proton, respectively.
The proton resonance width in this case is $0.45 \pm 0.10$ MeV \cite{skorodumov07}. The square of the ANC for the ground state obtained in Ref. \cite{banu09} is $
C_{l_{f}=1\,J_{f}= 3/2\,j= 1/2}^{2}= 2.53 \pm 0.30$ fm$^{-1}$.  This ANC was obtained in $jj$ coupling scheme and the last quantum number in the subscript $j=1/2$ is  the total angular momentum of the proton. However in the $R$-matrix method, the  $LS$  coupling scheme is used in which only the channel spin $I=1/2$ contributes, so the proton ANC of the ground state of ${}^{13}{\rm O}$ for the channel spin $I=1/2$  is $C_{l_{f}=1\,J_{f}=3/2\,I=1/2} =2/3\,C_{l_{f}=1\,J_{f}=3/2\, j=1/2}$  (see Ref.\cite{banu09}). For $R = 4.6$ fm, the channel radiative width amplitude $\gamma_{\gamma 3/2}^{1/2}(ch)$ for this case is $0.601+i~ 0.187$ eV$^{1/2}$. To calculate the internal radiative width amplitude, we use the same Woods-Saxon potential parameters as  in Ref. \cite{banu09}. The values of the potential depths $V_b$ and $V_c$ in this case are -45.15 MeV and -51.405 MeV, respectively. The obtained value of the single-particle ANC is $b_{1\,3/2\,1/2}= 2.16$ fm$^{-1}$, which for the given ANC  leads to the spectroscopic factor  $0.24$.  
Then for $\gamma_{\gamma 3/2}^{1/2}(int)$ we obtain $0.286$ eV$^{1/2}$. 

Now, using Eq. (\ref{a24}), we get for the total radiative width for this transition $\Gamma_{\gamma 3/2}^{1/2} =0.8 \pm 0.2$ eV, which changes only by  $4\%$ if we vary the channel radius by $6.5\%$.  The obtained value of the total radiative width  is close to the one reported in Ref. \cite{banu09}  for the channel radius $4.25$ fm, which is significantly larger than the value of  $24$ meV reported in Ref. \cite{wiescher89}. In fact, the radiative width of Ref. \cite{wiescher89} is even smaller than the lower limit $35$ meV of the radiative width obtained from the imaginary part of the channel radiative width amplitude  and has been questioned in Ref. \cite{akram_pang15}. Furthermore, the present value of the radiative width is smaller than the value of $3$ eV obtained in Ref. \cite{skorodumov07}, where a larger value of $1.85$ fm$^{-1}$ of the ANC for the channel spin $I=1/2$  was used.  

3. {\it Decay of} $^{17}$F($\frac{1}{2}^-$, $E_x=3.104$ {\it MeV}) $\rightarrow$ $^{17}$F($\frac{1}{2}^+$, $E_x=0.495$ {\it MeV}).\\  
This is an example of the resonance decay to the excited bound state. Here, the $1/2^-$ resonance at $E_{R}= 2.504$ MeV  of $^{17}$F decays to the $1/2^+$ bound state with the binding energy $\epsilon_{f} = 0.105$ MeV. The $1/2^+$ and $1/2^-$ states of $^{17}$F are obtained by coupling of the $^{16}$O ($0^+$) core with the
 2$s_{1/2}$ and 1$p_{1/2}$ protons, respectively. The square of the ANC for the $1/2^+$   state is $6490 \pm 680$ fm$^{-1}$ \cite{gagliardi99} and the partial resonance width of the proton is $19\pm1$ keV \cite{ajzenberg86}. From Eq. (\ref{a21}), using the channel radius $R= 4.9$ fm and the channel spin $I = 1/2$, we get $\gamma_{\gamma 1/2}^{1/2}(ch) = -0.202-i~0.179$ eV$^{1/2}$.  In this case  the experimental value of radiative width $(1.2\pm0.2) \times 10^{-2}$ eV \cite{rolfs73} is smaller than the lower limit imposed by the imaginary part of channel radiative width $3.16 \times 10^{-2}$ eV. 

In order to calculate the internal radiative width amplitude we use the same potential parameters as  in Ref. \cite{carlos03}. The potential depths $V_b$ and $V_c$ required for this case are -50.70 MeV and -20.98 MeV, respectively. The single-particle ANC obtained for the bound state is 79.145 fm$^{-1/2}$ and therefore the spectroscopic factor  is 1.04. Our calculated $\gamma_{\gamma 1/2}^{1/2}(int)$ for this transition is $ 0.15$ eV$^{1/2}$.  

The calculated total radiative width $\Gamma_{\gamma 1/2}^{1/2} = (3.5 \pm 0.6)  \times10^{-2}$ eV is contributed by the destructive interference of the internal and real part of the channel radiative width amplitudes and only is slightly higher than the radiative width $3.16 \times 10^{-2}$ eV  obtained from the imaginary part of the channel radiative width amplitude.   The total radiative width  changes by $14\%$ when the channel radius varies by  $6\%$.

4. {\it Decay of} $^{17}$F($\frac{5}{2}^-$, $E_x=3.857$ {\it MeV}) $\rightarrow$ $^{17}$F($\frac{5}{2}^+$, $E_x=0$ {\it MeV}).\\
 As a fourth example we consider the decay of the second resonance $5/2^-$ in $^{17}$F at $E_{R}= 3.257$ MeV  to the ground state $5/2^+$ with the binding energy  $0.6$ MeV. The $5/2^+$ and $5/2^-$ states of $^{17}$F are obtained by the coupling the $^{16}$O ($0^+$) core with 1$d_{5/2}$ and 1$f_{7/2}$ protons, respectively. The measured square of the proton ANC  for the ground state of ${}^{17}{\rm F}$  is $1.08 \pm 0.1$ fm$^{-1}$ \cite{gagliardi99} and the proton resonance width is  $ 1.5$ keV \cite{tilley93}. From our calculations we get $\gamma_{\gamma 5/2}^{5/2}(ch) = -0.049-i~0.0062$ eV$^{1/2}$ for $R = 4.9$ fm.
With the same potential parameters as  in the previous case, the potential depths $V_b$ and $V_c$ required for the ground and resonance states are $-53.45$ MeV and $-75.59$ MeV, respectively. The single-particle ANC of the ground state of ${}^{17}{\rm F}$  for the adopted bound-state potential is  $0.9313$ fm$^{-1/2}$, which corresponds to the spectroscopic factor $1.24$. 

Then the calculated internal radiative width amplitude is $-0.164$ eV$^{1/2}$. Thus in this case we obtain the constructive interference of the internal and external radiative width amplitudes when calculating the total radiative width, which is $\Gamma_{\gamma 5/2}^{5/2} =  0.046 \pm 0.005$ eV. 
Our calculated total radiative width is almost half of the value $0.11 \pm 0.02$ eV reported in Ref. \cite{tilley93}. The use of the upper limit of the ANC results in a $9\%$ increase  of our calculated $\Gamma_{\gamma 5/2}^{5/2}$  while $6\%$ variation of the channel radius  leads to the $4\%$ change of the total radiative width. 

5. {\it Decay of}  $^{12}$N($2^-$, $E_x=1.191$ {\it MeV}) $\rightarrow$ $^{12}$N($1^+$, $E_x=0$ {\it MeV}).\\
We now consider the decay of the resonance at  $E_{R}=0.591$ MeV  of $^{12}$N with the spin-parity $\,J_i = 2^-$ to the ground state $\,J_f = 1^+$ with the binding energy $\,0.6$ MeV. This transition contributes to the proton capture reaction $^{11}$C + $p$ $\rightarrow$ $^{12}$N, which is an important branching point in the alternative path from the slow $3\alpha$ process to produce CNO seed nuclei \cite{wiescher89,tang03}. In this case the $2^-$ and $1^+$ states of $^{12}$N are obtained by the coupling  of the $^{11}$C ($3/2^-$) core with 2$s_{1/2}$ and 1$p_{1/2}$ protons, respectively. The proton resonance width is $51 \pm 20$ keV \cite{sobtka13} and the measured square of the proton ANC for the ground state of ${}^{12}{\rm N}$ is $1.73 \pm 0.25$ fm$^{-1}$ \cite{tang03}. For the channel radius $R= 4.5$ fm and the channel spin $I = 2$, the channel reduced width amplitude  obtained using Eq. (\ref{a21}) is  $\gamma_{\gamma 1}^{2}(ch)=0.173 + 0.029$ eV$^{1/2}$. 
In order to calculate the bound-state wave function $\,\phi_{l_f=1\, J_f=1\, I=2}^B$ and $\,X_{int}$, we use the same Woods-Saxon parameters as  in case 1. The potential depths $V_b$ and $V_c$ are set  to  $-40.67$ MeV and $-55.18$ MeV, respectively. The calculated value of the internal reduced width amplitude is $\gamma_{\gamma 1}^{2}(int)= -0.101$ eV$^{1/2}$. It is important that the sign of this amplitude is negative what determines the destructive interference between the internal and the real part of the channel reduced width amplitudes
when calculating the total radiative width. 

Using Eq. (\ref{a24})  we get the total radiative width $\Gamma_{\gamma 1}^{2}  = (6.0 \pm 5.4) \times 10^{-3}$ eV for $R = 4.5$ fm. The calculated radiative width in this case, due to the destructive interference, is very sensitive to the choice of the channel radius: for the channel radius varying between $R = 4.2$ and 4.8 fm the total radiative width changes from  $\,\Gamma_{\gamma 1}^{2}  = 1.34 \times 10^{-2}$ to $1.36 \times 10^{-3}$ eV, respectively. In fact, radiative width for this case is a  controversial subject. 
The value of $ \Gamma_{\gamma 1}^{2}$ from the latest measurement at RIKEN is ($13 \pm 0.5) ~ \times 10^{-3}$ eV \cite{riken02}, whereas the previous GANIL measurement \cite{ganil95} gave $\Gamma_{\gamma 1}^{2}=6^{+7}_{-3.5} ~ \times 10^{-3}$ eV with quite large uncertainty.

6. {\it Decay of} $^{16}$O($1^-$, $E_x=12.44$ {\it MeV}) $\rightarrow$ $^{16}$O($0^+$, $E_x=0$ {\it MeV}).\\
As an another example, we consider the decay of the  $E_{R}=0.312$ MeV resonance of $^{16}$O with the spin-parity spin $J_{i}=1^-$ to the ground state $0^+$ with $\epsilon_{f}= 12.13$ MeV. We consider this example because of the importance of the reaction $^{15}{\rm N} + p \rightarrow {}^{16}{\rm O} + \gamma$, which provides a path from the CN cycle to the CNO bi-cycle and CNO tri-cycle. The cross section for this reaction is dominated by two $1^-$ resonances at  $0.312$ MeV and $0.962$ MeV \cite{rolfs74,hebbard60}. In this case the $1^-$ and $0^+$ states of $^{16}$O are obtained by coupling the $^{15}$N ($1/2^-$) core with 2$s_{1/2}$ and 1$p_{1/2}$ protons, respectively. The proton partial width $\Gamma_{J_{i}}^{(0)}$ is calculated from its reduced width amplitude $\gamma_{J_{i}}^{(0)}$ by using the relation, $\Gamma_{J_{i}}^{(0)} = 2 P_{l_i}(k_{R},\,R) [\gamma_{J_{i}}^{(0)}]^2$,  
where $[\gamma_{J_{i}}^{(0)}]^2 = 353.3$ keV and $R = 5.03$ fm are adopted from Ref. \cite{akram11}. The experimental  proton ANC for the ground state of ${}^{16}{\rm O}$ is $14.154$ fm$^{-1/2}$  Ref. \cite{akram11}. Our calculated channel radiative width amplitude  is $1.35 + i\, 0.0014$ eV$^{1/2}$. 

In order to calculate the internal part  of the radiative width amplitude in the potential model, we use the same Woods-Saxon parameters as  in Ref. \cite{akram11}. The potential depths $V_b$ and $V_c$ are set to -53.74 MeV and -48.99 MeV, respectively. The obtained single-particle ANC for the ground state of ${}^{16}{\rm O}$  is $10.314$ fm$^{-1/2}$, and the corresponding spectroscopic factor is $1.9$. Using Eqs. (\ref{a20})  and (\ref{a24}) we get $\gamma_{\gamma 0}^{1}(int) = - 3.90$ eV$^{1/2}$. 
Again, as in the previous case, we get the negative sign of the internal radiative width amplitude. Hence we have the destructive interference of the internal and real part of the channel radiative width when calculating the total radiative width, which is  $\Gamma_{\gamma 0}^{1}= 7.0 \pm 1.0$ eV. The present value overlaps with $\Gamma_{\gamma 0}^{1}=7.5$ eV obtained in Ref. \cite{akram11} using the $R$-matrix fit of the astrophysical factor for the $p+ {}^{15}{\rm N} \to {}^{16}{\rm O} + \gamma$ radiative capture. Our calculated radiative width is lower than $12 \pm 2$ eV quoted in \cite{tilley93} and overlaps with the low limit of $\Gamma_{\gamma 0}^{1}=9.5 \pm 1.7$ eV determined from the $^{12}$C + $\alpha$ resonance scattering \cite{tilley93}.  

7. {\it Decay of} $^{16}$O($1^-$, $E_x=13.090$ {\it MeV}) $\rightarrow$ $^{16}$O($0^+$, $E_x=0$ {\it MeV}). \\
Next we consider the decay of the second $1^{-}$ resonance of $^{16}$O at $E_{R}=0.962$ MeV to the ground state of $^{16}$O. The spin-parities of the initial and final states,     ground state potential, the single-particle ANC,  spectroscopic factor,  $\,R$ and ANC of the ground state are the same as those in the previous case. However, the  potential depth $V_c$  required to reproduce the resonance at  $E_{R}=0.962$ MeV is  $-9.92$ MeV. The squared partial reduced width amplitude for this second $1^-$ resonance is $(\gamma_{1}^{(0)})^2= 231.4$ keV \cite{akram11}.  Then our calculated  $\gamma_{\gamma 0}^{1}(ch)$ and $\gamma_{\gamma 0}^{1}(int)$ for 0.962 MeV resonance are $1.32 + i~0.101$ eV$^{1/2}$ and $-9.73$ eV$^{1/2}$, respectively. Again we observe a destructive interference of the internal and real part of the channel reduced width amplitudes when calculated the total radiative width for the decay of the second $1^{-}$ resonance of $^{16}$O, for which we obtain  $\Gamma_{\gamma 0}^{1}= 71 \pm 8.0$ eV. The present value is larger than $38.7$ eV \cite{leblanc10}, $44\pm8$ eV \cite{tilley93} but it lies between the values of $63.6$ eV \cite{akram11} and $88$ eV  \cite{hebbard60}.

8. {\it Decay of} $^{15}$O($3/2^+$, $E_x=6.79$ {\it MeV}) $\rightarrow$ $^{15}$O($1/2^-$, $E_x=0$ {\it MeV}).\\
One of the most interesting cases  is the decay of a subthreshold resonance. The subthreshold resonance is a weakly bound state (also called the subthreshold bound state) having its tail extended to the continuum which works like a resonance. The radiative capture to the ground state occurs as a capture to the subthreshold resonance at positive energy $E$ followed by its decay to the ground state by emitting the photon.
Here we consider the decay of the subthreshold resonance ($\,3/2^+$) in $\,^{15}$O with the binding energy $\epsilon_i=0.504$ MeV to the ground state $1/2^-$ of $^{15}$O with $\epsilon_{f} = 7.297$ MeV.  The value of the radiative  width  of this decay is one of the unsolved problems in the analysis  of the $^{14}$N + $p$  $\rightarrow$ $^{15}$O reaction, which is the bottleneck reaction of the CNO cycle \cite{rolfs88, akram03, adelberger11}. 
The $3/2^+$ and $1/2^-$ states of $^{15}$O are obtained by coupling of the $^{14}$N ($1^+$) core with 2$s_{1/2}$ and 1$p_{1/2}$ protons, respectively. It is clear from Eqs. (\ref{a28}) and (\ref{a21}), that the channel radiative width in this case is proportional to the product of the squares of the ANCs of these two bound state. For the channel spin $3/2$, the experimental squared ANCs of the ground and subthreshold states are $54\pm6.0$ fm$^{-1}$ and $24\pm5.0$ fm$^{-1}$ \cite{akram03}, respectively. 

With a channel radius of 5.5 fm we get $\gamma_{\gamma 1/2}^{3/2}(ch) = 0.83$ eV$^{1/2}$, which is real as the imaginary part in this case is negligible.  $\gamma_{\gamma 1/2}^{3/2}(int)$  is calculated by replacing the $X_{int}$ in Eq. (\ref{a20}) by the bound-state wave function corresponding to the subthreshold state, which is normalized to unity over the entire radial space ($0 \leq r < \infty$), multiplied by square root of its spectroscopic factor. Furthermore, $k_{\gamma}$ in this case is given by $(\epsilon_{i}-\epsilon_{f})/{\hbar c}$. The wave functions ($\phi_{l_{f}\,J_{f}\,I}^{B}$ and $X_{int}$) are generated by taking the potential parameters used in Ref. \cite{huang2010}. The potential depths and single-particle ANCs for the ground and subthreshold bound states from our adopted potentials are -43.45 MeV, -53.00 MeV, and 6.102 fm$^{-1/2}$, -5.75 fm$^{-1/2}$, respectively. From our calculations we get $\gamma_{\gamma 1/2}^{3/2}(int) = 2.98$ eV$^{1/2}$  and the total radiative width calculated using  Eq. (\ref{a24}) is $\Gamma_{\gamma 0}=14.5 \pm 3.5$ eV. This value is significantly larger than $0.4^{+0.34}_{-0.13}$ eV \cite{bertone01}, $0.95^{+0.6}_{-0.95}$ eV \cite{yamada04} and $0.85$ eV \cite{schurmann08} (lower limit only). The value of the radiative width obtained using the $R$-matrix fitting in Ref. \cite{akram_pang15} is 3.75 eV, which is also obtained for the constructive interference of the internal and external radiative width amplitudes. This shows that the potential model correctly predicts the sign of the internal part in this case but overestimates its magnitude.   

We summarize all our results in Table \ref{table:a1}, where we compare our calculated values of radiative width with those from previous measurements and theoretical estimates. 

\begin{table*}[ht]
\begin{center}
\caption{Calculated radiative width ($\Gamma_{\gamma}$) and its comparison with some of the previously measured or calculated values ($\Gamma_{\gamma}^M$). }
\label{table:a1}
\begin{tabular}{c|c|c|c}
\hline\hline
S.No. & Transition &  $\Gamma_{\gamma}$ (eV) & $\Gamma_{\gamma}^M$ (eV)   \\
               
\hline
1. & $^{13}$N($\frac{1}{2}^+, E_x=2.365$ MeV) $\rightarrow$ $^{13}$N($\frac{1}{2}^-, E_x=0$ MeV) &  0.61$\pm$0.06 & $0.65 \pm 0.07$ \cite{burtebaev}, 0.67 \cite{ajz70},  \\
&&& $0.50 \pm 0.04$ \cite{Ajzenberg91}, $0.45 \pm 0.05$ \cite{riess68}  \\
\hline
2. & $^{13}$O($\frac{1}{2}^+$, $E_x=2.69$ MeV) $\rightarrow$ $^{13}$O($\frac{3}{2}^-$, $E_x=0$ MeV) & 0.8$\pm$0.2 & 0.95 \cite{banu09}, 0.024 \cite{wiescher89},  \\ 
&&& 3 \cite{skorodumov07}, 1.12 \cite{akram_pang15}\\
\hline
3. & $^{17}$F($\frac{1}{2}^-$, $E_x=3.104$ MeV) $\rightarrow$ $^{17}$F($\frac{1}{2}^+$, $E_x=0.495$ MeV) & (3.5$\pm$0.6)$\times$10$^{-2}$ & (1.2 $\pm$ 0.2) $\times$ 10$^{-2}$ \cite{rolfs73} \\
\hline
4. & $^{17}$F($\frac{5}{2}^-$, $E_x=3.857$ MeV) $\rightarrow$ $^{17}$F($\frac{5}{2}^+$, $E_x=0$ MeV) &  0.046$\pm$0.005 & $0.11 \pm 0.02$ \cite{tilley93} \\
\hline
5. & $^{12}$N($2^-$, $E_x=1.191$ MeV) $\rightarrow$ $^{12}$N($1^+$, $E_x=0$ MeV) &  (6.0$\pm$5.4)$\times$10$^{-3}$ & (13 $\pm$ 0.5) $\times$ 10$^{-3}$ \cite{riken02},   \\
&&& $6^{+7}_{-3.5} ~ \times 10^{-3}$ \cite{ganil95} \\
\hline
6. & $^{16}$O($1^-$, $E_x=12.44$ MeV) $\rightarrow$ $^{16}$O($0^+$, $E_x=0$ MeV) &  $7.0 \pm 1.0$ & $ 12 \pm 2$ \cite{tilley93}, 7.5 \cite{akram11},  \\
&&& $9.5 \pm 1.7$ \cite{tilley93} \\
\hline
7. & $^{16}$O($1^-$, $E_x=13.090$ MeV) $\rightarrow$ $^{16}$O($0^+$, $E_x=0$ MeV) &  $71\pm8.0$  & 38.7 \cite{leblanc10}, $44\pm8$ \cite{tilley93},  \\
&&& 63.6 \cite{akram11}, 88 eV \cite{hebbard60}\\
\hline
8. & $^{15}$O($3/2^+$, $E_x=6.79$ MeV) $\rightarrow$ $^{15}$O($1/2^-$, $E_x=0$ MeV)&  $14.5\pm3.5$ & $>0.85$ \cite{schurmann08},  $0.4^{+0.34}_{-0.13}$ \cite{bertone01}, \\
&&&  $0.95^{+0.6}_{-0.95}$ eV \cite{yamada04}\\

\hline
\hline
\end{tabular}
\end{center}
\end{table*}

\section{Conclusions}
We have calculated the radiative width for the decay of a resonance to a bound state using the $R$-matrix formalism previously developed in \cite{holt78,
barker91}. However, instead of using internal radiative width as as fitting $R$-matrix parameter,  we applied a combined $R$-matrix formalism and 
potential model. The potential model was adopted to calculate the internal radiative width amplitude and its sign relative to the channel part.
The external part is determined by the ANCs and the proton resonance width. The total radiative width depends upon the type of the interference between the internal and external radiative width amplitudes.  
We apply our formalism to some cases of isolated resonance $\gamma$-decay for which single-level $R$-matrix is sufficient  and compare our calculations with some of the previous experimental or theoretical estimates. A consistent picture emerges for the relevance of interference of the internal and  external parts of the radiative widths.

\section*{Acknowledgment}
 A.M.M.  acknowledges support from the U.S. DOE grant numbers  DE-FG02-93ER40773 and DE-FG52-09NA29467 and  by the U.S. NSF grant number PHY-1415656. C.A.B. acknowledges support from the U.S. NSF Grant number 1415656 and the U.S. DOE Grant number DE-FG02-08ER41533.

\end{document}